\documentclass[12pt]{article}

\author{Yu.~M.~Zinoviev
       \thanks{E-mail address: Yurii.Zinoviev@ihep.ru} \\
        {\it Institute for High Energy Physics} \\
        {\it of National Research Center "Kurchatov Institute"} \\
        {\it Protvino, Moscow Region, 142280, Russia}}
\title{Massive higher spins in $d=3$ unfolded}

\date{}

\def\eptwo{\left\{ \phantom{|}^{\mu\nu}_{ab} \right\}}
\def\epthree{\varepsilon^{\mu\nu\alpha}}

\begin{document}

\maketitle

\begin{abstract}
In this paper we construct an unfolded formulation for the massive
bosonic higher spins in three dimensions as well as for their
partially massless limit of the maximal depth. We begin with the
equations for the one-forms coming from the frame-like gauge invariant
Lagrangians for such fields and then we supplement them with the
infinite number of equations for the appropriately chosen set of
zero-forms.
\end{abstract}

\thispagestyle{empty}
\newpage
\setcounter{page}{1}

\section*{Introduction}

In the recent paper \cite{BPSS15} new linear unfolded systems of
equations \cite{Vas91a,Vas94} (see also \cite{DS14} and references
therein) for the set of bosonic zero-forms in the three dimensional
anti-de Sitter space were investigated. They appear as one parameter
deformation for the linearized critical Prokushkin-Vasiliev theory
\cite{PV98} and were shown to be reach enough to describe
topologically massive higher spins, fractional spins and so on. In
this paper we are interested in the parity even massive bosonic higher
spins as well as the so called partially massless limits of the
maximal depth \cite{DW01a,Zin01,SV06,Zin08b}. Recall that in three
dimensions massless higher spins $s \ge 2$ as well as all partially
massless ones except the one with the maximal depth do not have any
physical degrees of freedom. At the same time, massive higher spins
have two physical degrees of freedom (exactly as massless one in
$d=4$), while the partially massless fields with the maximal depth
have just one. That is why we restrict ourselves with these two cases
here. 

In \cite{BPSS15} the authors begin directly with the new set of
equations for the zero-forms, then investigate their relation with the
representations of anti-de Sitter group $SO(2,2)$ and look for the
possibilities to supplement these zero-forms with the appropriate set
of gauge potentials. In the current paper we will go just the other
way round. Namely, we begin with the frame-like gauge invariant
description for the massive higher spins in $d=3$ \cite{BSZ12a,BSZ14a}
(for dimensions greater than three see \cite{Zin08b,PV10}) where from
the very beginning we perfectly know what these systems describe and
then we supplement the equations for the gauge one-forms with the
appropriate set of zero-forms and their equations. Note that exactly
as the gauge invariant description of massive higher spins itself our
construction works not only in anti-de Sitter space but in Minkowski
and de Sitter spaces as well provided $m^2 \ge (s-1)^2 \Lambda$.

The crucial questions for us was what is the correct set of zero-forms
one has to introduce to have unfolded formulation for massive
arbitrary spins. In $d \ge 4$ frame-like description for the massless
spin-$s$ field requires introduction of the physical, auxiliary and
extra one-forms:
$$
\Phi_\mu{}^{a(s-1)}, \quad \Omega_\mu{}^{a(s-1),b}, \quad
\Sigma_\mu{}^{a(s-1),b(k)}, \quad 2 \le k \le s-1
$$
while for the unfolded formulation one has to introduce a whole set of
zero-forms:
$$
W^{a(k),b(s)}, \quad s \le k
$$
where tensor $T^{a(k),b(l)}$ corresponds to the two-raw Young tableau
$Y(k,l)$. At the same time, unfolded descriptions for the spin-1 and
spin-0 fields requires, correspondingly:
$$
A_\mu, \quad F^{a(k),b}, \quad 1 \le k
$$
$$
\varphi, \quad \pi^{a(k)}, \quad 1 \le k
$$
The main idea of the gauge invariant description for massive higher
spin $s$ is that such description can be constructed out of the
massless fields with spins $s$, $s-1$, $\dots$, $0$. Similarly, in
\cite{PV10} the unfolded formulation for the massive higher spins in
$d \ge 4$ was constructed combining all the one-forms and zero-forms
necessary for the unfolding of all these massless fields. 

Now let us turn to the $d=3$. A lot of important consequences follow
from the simple fact that in $d=3$ antisymmetric second rank tensor is
equivalent to the vector:
$$
A^a = \varepsilon^{abc} B_{bc}
$$
In particular, one can show that any mixed tensor corresponding to the
Young tableau $Y(k,l)$ with $l \ge 2$ is identically zero. As a
result, frame-like formalism for the massless spin-$s$ fields with $s
\ge 2$ requires physical and auxiliary fields only, while all zero
forms are identically zero in agreement with the known fact that such
fields do not have any physical degrees of freedom. Thus only spin-1
and spin-0 fields require zero-forms for the unfolded description.
Moreover for the spin-1 case all zero-forms can be dualized into the
completely symmetric traceless tensors. This restricts us with the
just two sets of zero-forms, originating from the spin-1 and spin-0.
It was not at all evident that the very same zero-forms may describe
massive arbitrary spins but the results of \cite{BPSS15} show that it
is indeed possible.

Note that, in-principle, starting with the results of \cite{PV10},
setting most of the one-forms and zero-forms to zero, choosing
appropriate solution and scaling for the coefficients and dualizing
all the objects that are not completely symmetric in local indices,
one can obtain the desired unfolded description for $d=3$ case. But as
it often happens, especially when one deals with three-dimensional
theories, it is easier to straightforwardly derive such description
directly in $d=3$. Moreover, this allows us to take into account
peculiarities of three-dimensional theories that have no analogues in
$d \ge 4$.

The paper is organized as follows. In section 1 we begin with the
rather simple but instructive example of massive spin-2 and its
partially massless limit illustrating most of the features of our
construction. In section 2 we consider spin-3 (also both massive as
well as partially massless case) as one more concrete example
illustrating how the same set of zero-forms begins to describe higher
spin. At last in section 3 we gives the description for the massive
field with arbitrary spin. To make our paper to be self-contained we
supply three appendices which contains all necessary information on
the frame-like gauge invariant description of massive bosonic higher
spins (appendix A), its partially massless limits in-particular the
one with the maximal depth (appendix B) and also on the partial gauge
fixing (appendix C) that greatly simplifies calculations with massive
arbitrary spins.

\noindent
{\bf Notations and conventions.} We will work in the three dimensional
frame-like formalism where massless spin $s$ is described by the
physical $\Phi_\mu{}^{a_1a_1 \dots a_{s-1}}$ and auxiliary
$\Omega_\mu{}^{a_1a_2 \dots a_{s-1}}$ one-forms. Here Greek letters
denote world indices, while Latin letters denote local ones. All
objects will be assumed to be completely symmetric and traceless in
their local indices. To simplify formulas we will use the shorthand
notation such as
$$
\Phi_\mu{}^{a(k)} = \Phi_\mu{}^{a_1a_2 \dots a_k}
$$
Besides all the indices denoted by the same letter and placed on the
same level will be assumed symmetrized, e.g.
$$
e_\mu{}^a \eta^{a(k)} = e_\mu{}^{(a_1} \eta^{a_2 \dots a_{k+1})}
$$
where we use symmetrization without any normalization factor. We will
work in the (anti)-de Sitter space with arbitrary cosmological
constant
$\Lambda$, the frame field $e_\mu{}^a$ and the covariant derivative 
$D_\mu$ normalized so that
$$
D_{[\mu} D_{\nu]} \eta^a = - \Lambda e_{[\mu}{}^a \eta_{\nu]}
$$

\section{Spin 2}

In this section we consider a rather simple but instructive example of
spin 2, illustrating almost all the general features of our
construction. We begin with the partially massless case and then we
will turn to the general massive one.

\subsection{Partially massless case}

For the frame-like gauge invariant description of the partially
massless spin-2 we need two pairs of physical and auxiliary fields
($\Omega_\mu{}^a$, $\Phi_\mu{}^a$) and 
($B^a$, $A_\mu$).\footnote{Recall that in $d=4$ such field has four
physical degrees of freedom, namely helicities $\pm 2,\pm 1$ (and this
explains the set of fields introduced), while in $d=3$ it has only
one.} The free Lagrangian describing partially massless spin-2 in the
three dimensional de Sitter space with positive cosmological term
$\Lambda$ has the form:
\begin{eqnarray}
{\cal L}_0 &=& \frac{1}{2} \eptwo \Omega_\mu{}^a \Omega_\nu{}^b -
\epthree \Omega_\mu{}^a D_\nu \Phi_{\alpha,a} + B^a
B_a - \epthree B_\mu D_\nu A_\alpha \nonumber \\
 && - m \epthree  [\Omega_{\mu,\nu} A_\alpha - B_\mu \Phi_{\nu,\alpha}
]
\end{eqnarray}
where $\eptwo = e^\mu{}_a e^\nu{}_b - e^\nu{}_a e^\mu{}_b$ and $m^2 =
\Lambda$. This Lagrangian is invariant under the following gauge
transformations:
\begin{eqnarray}
\delta \Omega_\mu{}^a &=& D_\mu \eta^a, \qquad
\delta \Phi_\mu{}^a = D_\mu \xi^a  + \varepsilon_\mu{}^{ab} \eta_b + m
e_\mu{}^a \xi \nonumber \\
\delta B^a &=& - m \eta^a, \qquad
\delta A_\mu = D_\mu \xi + m \xi_\mu
\end{eqnarray}
There exist four gauge invariant objects:
\begin{eqnarray}
{\cal F}_{\mu\nu}{}^a &=& D_{[\mu} \Omega_{\nu]}{}^a - m e_{[\mu}{}^a
B_{\nu]} \nonumber \\
{\cal T}_{\mu\nu}{}^a &=& D_{[\mu} \Phi_{\nu]}{}^a +
\varepsilon_{[\mu}{}^{ab} \Omega_{\nu]b} + m e_{[\mu}{}^a A_{\nu]}
\nonumber \\
{\cal B}_\mu{}^a &=& D_\mu B^a + m \Omega_\mu{}^a \\
{\cal A}_{\mu\nu} &=& D_{[\mu} A_{\nu]} - 2 \varepsilon_{\mu\nu a}
B^a - m \Phi_{[\mu,\nu]} \nonumber
\end{eqnarray}
Let us take the first four unfolded equations in the form:
\begin{eqnarray}
0 &=& D_{[\mu} \Omega_{\nu]}{}^a - m e_{[\mu}{}^a B_{\nu]} \nonumber 
\\
0 &=& D_{[\mu} \Phi_{\nu]}{}^a + \varepsilon_{[\mu}{}^{ab} 
\Omega_{\nu]b} + m e_{[\mu}{}^a A_{\nu]} \nonumber \\
0 &=& D_\mu B^a + m \Omega_\mu{}^a - B_\mu{}^a \\
0 &=& D_{[\mu} A_{\nu]} - 2 \varepsilon_{\mu\nu a} B^a - m 
\Phi_{[\mu,\nu]} \nonumber
\end{eqnarray}
where the zero-form $B^{ab}$ is symmetric and traceless.\footnote{
Note once again that in $d=3$ massless fields with spins $s \ge 2$ do
not have any physical degrees of freedom, thus there is no need to
introduce zero forms for them.} All these equations except the third
one are already consistent while the consistency for the third one
gives:
\begin{eqnarray}
0 &=& D_{[\mu} D_{\nu]} B^a + m D_{[\mu} \Omega_{\nu]}{}^a - D_{[\mu}
B_{\nu]}{}^a \nonumber \\
 &=& - \Lambda e_{[\mu}{}^a B_{\nu]} + m^2 e_{[\mu}{}^a B_{\nu]} -
D_{[\mu} B_{\nu]}{}^a \nonumber \\
 &=& D_{[\mu} B_{\nu]}{}^a
\end{eqnarray}
Taking into account that we deal with the parity even theory, we
choose the following ansatz for the remaining equations ($k \ge 2$):
\begin{equation}
0 = D_\mu B^{a(k)} - B_\mu{}^{a(k)} + E_k [ e_\mu{}^a B^{a(k-1)}
- \frac{2}{(2k-1)} g^{a(2)} B_\mu{}^{a(k-2)} ], \quad B_2 = 0
\end{equation}
where all $B^{a(k)}$ are symmetric, traceless and gauge invariant.
Consistency for with ansatz requires:
\begin{equation}
\frac{(2k+3)}{(2k+1)} E_{k+1} - E_k - \Lambda = 0
\end{equation}
and we obtain the solution:
\begin{equation}
E_k = \frac{(k^2-4)}{(2k+1)} \Lambda
\end{equation}

\subsection{General massive case}

The frame-like gauge invariant description for the general massive
spin-2 requires also a pair ($\pi^a$, $\varphi$). The free Lagrangian
looks like:
\begin{eqnarray}
{\cal L}_0 &=& \frac{1}{2} \eptwo \Omega_\mu{}^a \Omega_\nu{}^b -
\epthree \Omega_\mu{}^a D_\nu \Phi_{\alpha,a} + B_a{}^2 - \epthree 
B_\mu D_\nu A_\alpha  - \pi_a{}^2 + \pi^\mu D_\mu \varphi  \nonumber\\
&& - m \epthree [ \Omega_{\mu\nu} A_\alpha - B_\mu \Phi_{\nu\alpha}] +
2M  \pi^\mu A_\mu \nonumber \\
 && + \frac{M^2}{2} \eptwo \Phi_\mu{}^a \Phi_\nu{}^b + mM \Phi \varphi
+ \frac{3m^2}{4} \varphi^2
\end{eqnarray}
where $M^2 = m^2 - \Lambda$. Note that such description works in
anti-De Sitter, Minkowski and De Sitter space provided $m^2 \ge
\Lambda$. This Lagrangian is invariant under the following gauge
transformations:
\begin{eqnarray}
\delta \Omega_\mu{}^a &=& D_\mu \eta^a + M^2
\varepsilon_\mu{}^{ab} \xi_b \nonumber \\
\delta \Phi_\mu{}^a &=& D_\mu \xi^a + \varepsilon_\mu{}^{ab}
\eta_b + m e_\mu{}^a \xi \nonumber \\ 
\delta A_\mu &=& D_\mu \xi + m \xi_\mu, \qquad
\delta B^a = - m \eta^a \\
\delta \varphi &=& - 2M  \xi, \qquad
\delta \pi^a = mM \xi^a \nonumber
\end{eqnarray}
Correspondingly, in this case one can construct six gauge invariant
objects:
\begin{eqnarray}
{\cal F}_{\mu\nu}{}^a &=& D_{[\mu} \Omega_{\nu]}{}^a - m
e_{[\mu}{}^a B_{\nu]} + M^2 \varepsilon_{[\mu}{}^{ab} 
\Phi_{\nu],b} - Mm \varepsilon_{\mu\nu}{}^a \varphi \nonumber \\
{\cal T}_{\mu\nu}{}^a &=& D_{[\mu} \Phi_{\nu]}{}^a + 
\varepsilon_{[\mu}{}^{ab} \Omega_{\nu],b} + m e_{[\mu}{}^a
A_{\nu]}  \nonumber  \\
{\cal B}_\mu{}^a &=& D_\mu B^a + m \Omega_\mu{}^a - M
\varepsilon_\mu{}^{ab} \pi_b \nonumber   \\
{\cal A}_{\mu\nu} &=& D_{[\mu} A_{\nu]} - 2 
\varepsilon_{\mu\nu a} B^a - m \Phi_{[\mu,\nu]}   \\
\Pi_\mu{}^a &=& D_\mu \pi^a - M \varepsilon_\mu{}^{ab} B_b
- Mm \Phi_\mu{}^a - \frac{m^2}{2} e_\mu{}^a \varphi \nonumber  \\
\Phi_\mu &=& D_\mu \varphi - 2 \pi_\mu + 2M A_\mu \nonumber
\end{eqnarray}
Thus for the first six unfolded equations we take
\begin{equation}
{\cal F}_{\mu\nu}{}^a = 0, \qquad {\cal T}_{\mu\nu}{}^a = 0, \qquad
{\cal A}_{\mu\nu} = 0, \qquad \Phi_\mu = 0
\end{equation}
which are already consistent as well as:
\begin{eqnarray}
0 &=& D_\mu B^a + m \Omega_\mu{}^a - M \varepsilon_\mu{}^{ab} \pi_b -
B_\mu{}^a \nonumber \\
0 &=& D_\mu \pi^a - M \varepsilon_\mu{}^{ab} B_b - Mm \Phi_\mu{}^a -
\frac{m^2}{2} e_\mu{}^a \varphi - \pi_\mu{}^a
\end{eqnarray}
where the zero-forms $B^{ab}$ and $\pi^{ab}$ are symmetric and
traceless.\footnote{Note that for the spin-2 case the frame-like
formalism in $d=3$ requires essentially the same (up to dualization)
set of fields as in $d \ge 4$, the main difference being the absence
of the Weyl zero-form. As a result the structure of equations 
(11)-(13) is the same as that of equations (3.53)-(3.58) in
\cite{PV10}.}
Their consistency requires:
\begin{eqnarray}
D_{[\mu} B_{\nu]}{}^a = M \varepsilon_{[\mu}{}^{ab} \pi_{\nu],b} 
\nonumber \\
D_{[\mu} \pi_{\nu]}{}^a = M \varepsilon_{[\mu}{}^{ab} B_{\nu],b}
\end{eqnarray}
So we introduce the following general ansatz for the remaining
equations ($k \ge 2$):\footnote{This ansatz is essentially the same as
in \cite{BPSS15}. Note that the zero-forms $B$ and $\pi$ here have
opposite parities, so this ansatz does not break parity. Recall that
the whole family of the models with matrix coefficients $\mu$ given in
\cite{BPSS15} contains both parity odd and parity even ones. Note also
that seemingly difference in some coefficients is related with
different conventions on symmetrization.}
\begin{eqnarray}
0 &=& D_\mu B^{a(k)} - B_\mu{}^{a(k)} + A_k \varepsilon_{\mu b}{}^a
\pi^{a(k-1)b} \nonumber \\
 && + E_k [ e_\mu{}^a B^{a(k-1)} - \frac{2}{(2k-1)} g^{a(2)}
B_\mu{}^{a(k-2)} ] \nonumber \\
0 &=& D_\mu \pi^{a(k)} - \pi_\mu{}^{a(k)} + C_k 
\varepsilon_{\mu b}{}^a B^{a(k-1)b} \\
 && + D_k [ e_\mu{}^a \pi^{a(k-1)} - \frac{2}{(2k-1)} g^{a(2)} 
\pi_\mu{}^{a(k-2)} ] \nonumber
\end{eqnarray}
where
$$
A_2 = C_2 = \frac{M}{3}, \qquad E_2 = D_2 = 0
$$
Consistency for these equations requires:
\begin{equation}
A_k = C_k, \qquad E_k = D_k, \qquad
A_{k+1} = \frac{k}{(k+2)} A_k
\end{equation}
\begin{equation}
B_{k+1} = \frac{(2k+1)}{(2k+3)} [ B_k + A_k{}^2 + \Lambda ]
\end{equation}
These relations can be easily solved and give the solution:
\begin{equation}
A_k = \frac{2M}{k(k+1)}, \qquad
E_k = \frac{(k^2-4)}{(2k+1)} [ \frac{M^2}{k^2} + \Lambda ]
\end{equation}

\subsection{Partial gauge fixing}

It is rather well known \cite{AT86,Wit88} (see also \cite{Gom13} and
references therein) that for the massless higher spin fields in three
dimensional anti-de Sitter space one can use a separation of variables
that greatly simplifies all calculations. In this subsection we will
show that such separation is possible for the massive spin-2 as well
(for an arbitrary spin case see appendix C) provided one uses partial
gauge fixing removing the scalar field. Moreover, such procedure works
not only in anti-de Sitter, but in Minkowski and de Sitter spaces as
well, provided $m^2 > \Lambda$.

Let us partially fix the gauge setting $\varphi = 0$, solve the
corresponding constraint $\Phi_\mu = 0 \Rightarrow A_\mu = \pi_\mu/M$
and change the normalization $\pi^a \Rightarrow M \pi^a$ (because now
this field will play the role of physical field and not that of the
auxiliary one). The resulting Lagrangian will take the form:
\begin{eqnarray}
{\cal L} &=& \frac{1}{2} \eptwo \Omega_\mu{}^a \Omega_\nu{}^b -
\varepsilon^{\mu\nu\alpha} \Omega_\mu{}^a D_\nu \Phi_{\alpha,a} + B^a
B_a - \varepsilon^{\mu\nu\alpha} B_\mu D_\nu \pi_\alpha \nonumber \\
 && - m \varepsilon^{\mu\nu\alpha} \Omega_{\mu,\nu} \pi_\alpha + m
\varepsilon^{\mu\nu\alpha} B_\mu \Phi_{\nu,\alpha} + \frac{M^2}{2}
\eptwo \Phi_\mu{}^a \Phi_\nu{}^b + M^2 \pi^a \pi_a
\end{eqnarray}
This Lagrangian is still invariant under the two remaining gauge
transformations:
\begin{eqnarray}
\delta \Omega_\mu{}^a &=& D_\mu \eta^a + M^2 \varepsilon_\mu{}^{ab}
\xi_b \nonumber \\
\delta \Phi_\mu{}^a &=& D_\mu \xi^a + \varepsilon_\mu{}^{ab} \eta_b \\
\delta B^a &=& - m\eta^a, \qquad \delta \pi^a = m\xi^a \nonumber
\end{eqnarray}
Let us introduce the new variables:
$$
\hat{\Omega}_\mu{}^a = \Omega_\mu{}^a + M \Phi_\mu{}^a, \qquad
\hat{\Phi}_\mu{}^a = \Omega_\mu{}^a - M \Phi_\mu{}^a
$$
$$
\hat{B}^a = B^a - M \pi^a, \qquad
\hat{\pi}^a = B^a + M \pi^a
$$
Then the Lagrangian decompose into two independent parts:
$$
{\cal L} = \frac{1}{4M} [ {\cal L}(\hat{\Omega},\hat{B}) - 
{\cal L}(\hat{\Phi},\hat{\pi}) ]
$$
where, for example,
\begin{eqnarray}
{\cal L}(\hat{\Omega},\hat{B}) &=& 2M \eptwo \hat{\Omega}_\mu{}^a
\hat{\Omega}_\nu{}^b - \epthree \hat{\Omega}_\mu{}^a D_\nu
\hat{\Omega}_{\alpha.a} + 2M \hat{B}^a \hat{B}_a + 
\epthree \hat{B}_\mu D_\nu \hat{B}_\alpha \nonumber \\
 && + 2m \epthree \hat{\Omega}_{\mu,\nu} \hat{B}_\alpha
\end{eqnarray}
This Lagrangian has only one gauge symmetry, namely:
\begin{equation}
\delta \hat{\Omega}_\mu{}^a = D_\mu \hat{\eta}^a + M 
\varepsilon_\mu{}^{ab} \hat{\eta}_b, \qquad
\delta \hat{B}^a = - m \hat{\eta}^a, \qquad
\hat{\eta}^a = \eta^a + M \xi^a
\end{equation}
Correspondingly, there exist two gauge invariant objects:
\begin{eqnarray}
\hat{\cal F}_{\mu\nu}{}^a &=& D_{[\mu} \hat{\Omega}_{\nu]}{}^a - m
e_{[\mu}{}^a \hat{B}_{\nu]} + M \varepsilon_{[\mu}{}^{ab}
\hat{\Omega}_{\nu],b} \nonumber \\
\hat{\cal B}_\mu{}^a &=& D_\mu \hat{B}^a + m \hat{\Omega}_\mu{}^a + M
\varepsilon_\mu{}^{ab} \hat{B}_b
\end{eqnarray}
As the first two unfolded equations we take:
\begin{eqnarray}
0 &=& D_{[\mu} \hat{\Omega}_{\nu]} - m e_{[\mu} \hat{B}_{\nu]} + M
\varepsilon_{[\mu}{}^{ab} \hat{\Omega}_{\nu],b} \nonumber \\
0 &=& D_\mu \hat{B}^a + m \hat{\Omega}_\mu{}^a + M 
\varepsilon_\mu{}^{ab} \hat{B}_b - \hat{B}_\mu{}^a
\end{eqnarray}
The first one appears to be consistent, while the consistency of the
second one requires:
\begin{equation}
D_{[\mu} \hat{B}_{\nu]}{}^a = - M \varepsilon_{[\mu}{}^{ab} 
\hat{B}_{\nu],b}
\end{equation}
So we choose the following ansatz for the remaining equations ($k \ge
2$):\footnote{Note that these equations alone do break parity. The
original parity even theory will be restored if we combine both pairs
($\hat{\Omega}$, $\hat{B}$) and ($\hat{\Phi}$, $\hat{\pi}$) with the
correct coefficients.}
\begin{eqnarray}
0 &=& D_\mu \hat{B}^{a(k)} - \hat{B}_\mu{}^{a(k)} - A_k 
\varepsilon_{\mu b}{}^a \hat{B}^{a(k-1)b} \nonumber \\
 && + E_k [ e_\mu{}^a \hat{B}^{a(k-1)} - \frac{2}{(2k-1)} g^{a(2)}
\hat{B}_\mu{}^{a(k-2)} ]
\end{eqnarray}
where
$$
A_2 = \frac{M}{3}, \qquad E_2 = 0
$$
The consistency for these equations leads us to the following
solution:
\begin{equation}
A_k = \frac{2M}{k(k+1)}, \qquad
E_k = \frac{(k^2-4)}{(2k+1)} [ \frac{M^2}{k^2} + \Lambda ]
\end{equation}

\section{Spin 3}

In this section we consider one more concrete example that will
illustrate how the very same set of zero forms can describe higher
spins. Again we begin with the partially massless case (of maximal
depth) and then we will turn to the general massive one.

\subsection{Partially massless case}

The frame-like gauge invariant description requires three pairs of
physical and auxiliary fields ($\Omega_\mu{}^{ab}$, 
$\Phi_\mu{}^{ab}$), ($\Omega_\mu{}^a$, $\Phi_\mu{}^a$) and 
($B^a$, $A_\mu$). The free Lagrangian has the form:
\begin{eqnarray}
{\cal L}_0 &=& - \eptwo \Omega_\mu{}^{ac} \Omega_\nu{}^b{}_c +
\epthree \Omega_\mu{}^{ab} D_\nu \Phi_{\alpha,ab} + \frac{1}{2}
\eptwo \Omega_\mu{}^a \Omega_\nu{}^b \nonumber \\
 && - \epthree \Omega_\mu{}^a D_\nu \Phi_{\alpha,a} + \frac{1}{2} B^a
B_a - \epthree B_\mu D_\nu A_\alpha \nonumber \\
 && - \epthree [ 3b_2 \Omega_{\mu,\nu}{}^a  \Phi_{\alpha,a} + b_2
\Phi_{\mu,\nu}{}^a \Omega_\alpha{}^a - 2b_1 \Omega_{\mu,\nu} A_\alpha
+ b_1 \Phi_{\mu,\nu} B_\alpha ]
\end{eqnarray}
where
$$
b_2{}^2 = \frac{2\Lambda}{3}, \qquad b_1{}^2 = \frac{4\Lambda}{3}
$$
This Lagrangian is invariant under the following gauge
transformations:
\begin{eqnarray}
\delta \Omega_\mu{}^{ab} &=& D_\mu \eta^{ab} - \frac{b_2}{2} 
( e_\mu{}^{(a} \eta^{b)} - \frac{2}{3} g^{ab} \eta_\mu) \nonumber \\
\delta \Phi_\mu{}^{ab} &=& D_\mu \xi^{ab} - \varepsilon_\mu{}^{c(a}
\eta^{b)}{}_c - \frac{3b_2}{2} ( e_\mu{}^{(a} \xi^{b)} - \frac{2}{3}
g^{ab} \xi_\mu) \nonumber \\
\delta \Omega_\mu{}^a &=& D_\mu \eta^a - 3b_2 \eta_\mu{}^a \\
\delta \Phi_\mu{}^a &=& D_\mu \xi^a + \varepsilon_\mu{}^{ab} \eta_b -
b_2 \xi_\mu{}^a + 2b_1 e_\mu{}^a \xi \nonumber \\
\delta B^a &=& - 2b_1 \eta^a, \qquad
\delta A_\mu = D_\mu \xi + b_1 \xi_\mu \nonumber
\end{eqnarray}
Correspondingly, there exist six gauge invariant objects and hence the
first six unfolded equations:
\begin{eqnarray}
0 &=& D_{[\mu} \Omega_{\nu]}{}^{ab} - \frac{b_2}{2} ( e_{[\mu}{}^{(a}
\Omega_{\nu]}{}^{b)} + \frac{2}{3} g^{ab} \Omega_{[\mu,\nu]})
\nonumber \\
0 &=& D_{[\mu} \Phi_{\nu]}{}^{ab} - \varepsilon_{[\mu}{}^{c(a}
\Omega_{\nu]}{}^{b)}{}_c - \frac{3b_2}{2} ( e_{[\mu}{}^{(a} 
\Phi_{\nu]}{}^{b)} + \frac{2}{3} g^{ab} \Phi_{[\mu,\nu]}) \nonumber \\
0 &=& D_{[\mu} \Omega_{\nu]}{}^a + 3b_2 \Omega_{[\mu,\nu]}{}^a -
b_1 e_{[\mu}{}^a B_{\nu]} \nonumber \\
0 &=& D_{[\mu} \Phi_{\nu]}{}^a + \varepsilon_{[\mu}{}^{ab} 
\Omega_{\nu],b} + b_2 \Phi_{[\mu,\nu]}{}^a + 2b_1 e_{[\mu}{}^a
A_{\nu]} \\
0 &=& D_\mu B^a + 2b_1 \Omega_\mu{}^a - B_\mu{}^a \nonumber \\
0 &=& D_{[\mu} A_{\nu]} - \varepsilon_{\mu\nu a} B^a - b_1
\Phi_{[\mu,\nu]} \nonumber
\end{eqnarray}
where gauge invariance requires that
\begin{equation}
\delta B^{ab} = - 6b_2b_1 \eta^{ab}
\end{equation}
All equations except the one for the $B^a$ are already consistent, so
we have to deal with the one equation. It looks exactly the same as in
the spin-2 case considered above, but the crucial difference is that
now the zero-form $B^{ab}$ is not gauge invariant. Indeed, consistency
for the $B^a$ equation gives now:
\begin{equation}
D_{[\mu} B_{\nu]}{}^a = - 6b_2b_1 \Omega_{[\mu,\nu]}{}^a + 
\frac{5b_2{}^2}{2} e_{[\mu}{}^a B_{\nu]}
\end{equation}
So we take the following form for the next equation:
\begin{equation}
0 = D_\mu B^{ab} + 6b_2b_1 \Omega_\mu{}^{ab} - \frac{3b_2{}^2}{2} 
( e_\mu{}^{(a} B^{b)} - \frac{2}{3} g^{ab} B_\mu ) - B_\mu{}^{ab}
\end{equation}
In turn, its consistency leads to
\begin{equation}
D_{[\mu} B_{\nu]}{}^{a(2)} = 0
\end{equation}
where $B^{a(3)}$ (as well as all $B^{a(k)}$, $k \ge 3$) are gauge
invariant. Thus, taking into account that we have parity even theory,
we obtain the following equations for all higher rank zero-forms
($k \ge 3$):
\begin{equation}
0 = D_\mu B^{a(k)} - B_\mu{}^{a(k)} + \frac{(k^2-9)}{(2k+1)} \Lambda [
e_\mu{}^a B^{a(k-1)} - \frac{2}{(2k-1)} g^{a(2)} B_\mu{}^{a(k-2)} ]
\end{equation}

\subsection{General massive case}

This time to simplify presentation from the very beginning we will use
the possibility to separate the variables after partial gauge fixing
(see appendix C) and consider the subsystem containing the fields
($\hat{\Omega}_\mu{}^{ab}$, $\hat{\Omega}_\mu{}^a$, $\hat{B}^a$) only.
The corresponding Lagrangian looks like:
\begin{eqnarray}
{\cal L} &=& - \frac{1}{4M_2} [ 2M_2 \eptwo \hat{\Omega}_\mu{}^{ac}
\hat{\Omega}_\nu{}^b{}_c - \epthree \hat{\Omega}_\mu{}^{a(2)} D_\nu
\hat{\Omega}_{\alpha,a(2)} ] \nonumber \\
 && + \frac{1}{4M_1} [ \eptwo \hat{\Omega}_\mu{}^a 
\hat{\Omega}_\nu{}^b - \epthree \hat{\Omega}_\mu{}^a D_\nu
\hat{\Omega}_{\alpha,a} ] \nonumber \\
 && + \frac{1}{4M_1} [ M_1 \hat{B}^a \hat{B}_a +
\epthree \hat{B}_\mu D_\nu \hat{B}_\alpha ] \nonumber \\
 && + \varepsilon^{\mu\nu\alpha} [ - \frac{b_2}{2M_2}
\hat{\Omega}_{\mu,\nu}{}^a \hat{\Omega}_{\alpha,a} + \frac{b_1}{2M_1}
\hat{\Omega}_{\mu,\nu} \hat{B}_\alpha ]
\end{eqnarray}
where
$$
M_2{}^2 = \frac{1}{4} [ m^2 - 4\Lambda ], \qquad
M_1{}^2 = \frac{9}{4} [ m^2 - 4\Lambda ] 
$$
$$
b_2{}^2 = \frac{m^2}{6}, \qquad 
b_1{}^2 = \frac{4}{3} [ m^2 - 3\Lambda ]
$$
This Lagrangian is invariant under the following gauge
transformations:
\begin{eqnarray}
\delta \hat{\Omega}_\mu{}^{a(2)} &=& D_\mu \hat{\eta}^{a(2)} -
\frac{b_2}{2} [ e_\mu{}^a \hat{\eta}^a - \frac{2}{3} g^{a(2)}
\hat{\eta}_\mu ] - M_2 \varepsilon_{\mu b}{}^a \hat{\eta}^{ab}
\nonumber \\
\delta \hat{\Omega}_\mu{}^a &=& D_\mu \hat{\eta}^a - 3b_2
\hat{\eta}_\mu{}^a + M_1 \varepsilon_\mu{}^{ab} \hat{\eta}_b \\
\delta \hat{B}^a &=& - 2b_1 \hat{\eta}^a \nonumber
\end{eqnarray}
There exist three gauge invariant objects giving us the first three
unfolded equations:
\begin{eqnarray}
0 &=& D_{[\mu} \hat{\Omega}_{\nu]}{}^{a(2)} - \frac{b_2}{2} 
[ e_{[\mu}{}^a \hat{\Omega}_{\nu]}{}^a + \frac{2}{3} g^{a(2)} 
\hat{\Omega}_{[\mu,\nu]} ] - M_2 \varepsilon_{[\mu b}{}^a
\hat{\Omega}_{\nu]}{}^{ab} \nonumber \\
0 &=& D_{[\mu} \hat{\Omega}_{\nu]}{}^a + 3b_2
\hat{\Omega}_{[\mu,\nu]}{}^a + M_1 \varepsilon_{[\mu}{}^{ab}
\hat{\Omega}_{\nu],b} - b_1 e_{[\mu}{}^a \hat{B}_{\nu]} \\
0 &=& D_\mu \hat{B}^a + 2b_1 \hat{\Omega}_\mu{}^a + M_1 
\varepsilon_\mu{}^{ab} \hat{B}_b - \hat{B}_\mu{}^a \nonumber
\end{eqnarray}
where
$$
\delta \hat{B}^{a(2)} = - 6b_2b_1 \hat{\eta}^{a(2)}
$$
The first two are consistent, while for the third one we obtain:
\begin{equation}
D_{[\mu} B_{\nu]}{}^a = - 6b_2b_1 \hat{\Omega}_{[\mu,\nu]}{}^a - M_1
\varepsilon_{[\mu}{}^{ab} \hat{B}_{\nu],b} + \frac{5b_2{}^2}{2}
e_{[\mu}{}^a \hat{B}_{\nu]}
\end{equation}
So we take the following form for the next equation:
\begin{equation}
0 = D_\mu \hat{B}^{a(2)} + 6b_2b_1 \hat{\Omega}_\mu{}^{a(2)} - M_2
\varepsilon_{\mu b}{}^a \hat{B}^{ab} - \frac{3b_2{}^2}{2}
[ e_\mu{}^a \hat{B}^a - \frac{2}{3} g^{a(2)} \hat{B}_\mu ] - 
\hat{B}_\mu{}^{a(2)}
\end{equation}
In turn, the consistency for the last equations gives
\begin{equation}
D_{[\mu} \hat{B}_{\nu]}{}^{a(2)} = M_2 \varepsilon_{[\mu b}{}^a
\hat{B}_{\nu]}{}^a
\end{equation}
Taking into account that all $\hat{B}^{a(k)}$, $k \ge 3$ are gauge
invariant, we take the following ansatz for the remaining equations:
\begin{eqnarray}
0 &=& D_\mu \hat{B}^{a(k)} - \hat{B}_\mu{}^{a(k)} - A_k 
\varepsilon_{\mu b}{}^a \hat{B}^{a(k-1)b} \nonumber \\
 && + E_k [ e_\mu{}^a \hat{B}^{a(k-1)} - \frac{2}{(2k-1)} g^{a(2)}
\hat{B}_\mu{}^{a(k-2)} ]
\end{eqnarray}
where
$$
A_3 = \frac{M_2}{2}, \qquad E_3 = 0
$$
The consistency of these equations leads to the following solution:
\begin{equation}
A_k = \frac{6M_2}{k(k+1)}, \qquad
E_k = \frac{(k^2-9)}{(2k+1)} [ \frac{4M_2{}^2}{k^2} + \Lambda ]
\end{equation}

\section{Arbitrary spin}

Now we are ready to consider generalization to the case of arbitrary
spin. Once again, a separate subsection will be devoted to the
partially massless case of the maximal depth.

\subsection{Partially massless case}

The Lagrangian, gauge transformations and the whole set of gauge
invariant objects are given in the appendix B. From these formulas one
can see that auxiliary fields $\Omega_\mu{}^{a(k)}$ and $B^a$ generate
a closed subsystem in a sense that they transform non-trivially under
the $\eta$ transformations only and as a result their gauge invariant
objects contain only the auxiliary fields themselves. Thus we begin
with equations for the auxiliary fields (one can easily check that the
equations for the physical ones are consistent):
\begin{eqnarray}
0 &=& D_{[\mu} \Omega_{\nu]}{}^{a(k)} + \frac{(k+2)b_{k+1}}{k} 
\Omega_{[\mu,\nu]}{}^{a(k)} \nonumber \\
 && - \frac{b_k}{k} [ e_{[\mu}{}^a \Omega_{\nu]}{}^{a(k-2)} + 
\frac{2}{(2k-1)} g^{a(2)} \Omega_{[\mu,\nu]}{}^{a(k-2)} ] \nonumber \\
0 &=& D_{[\mu} \Omega_{\nu]}{}^a + 3b_2 \Omega_{[\mu,\nu]}{}^a -
2b_1{}^2 e_{[\mu}{}^a B_{\nu]} \\
0 &=& D_\mu B^a + \Omega_\mu{}^a - B_\mu{}^a \nonumber
\end{eqnarray}
where parameters $b_k$ are given in (\ref{b_param}) and to simplify
subsequent formulas we have changed normalization for the zero forms.
We will also need the gauge transformations that look like:
\begin{eqnarray}
\delta \Omega_\mu{}^{a(k)} &=& D_\mu \eta^{a(k)} - 
\frac{(k+2)b_{k+1}}{k} \eta_\mu{}^{a(k)} \nonumber \\
 && - \frac{b_k}{k} ( e_\mu{}^a \eta^{a(k-1)} - \frac{2}{(2k-1)}
g^{a(2)} \eta_\mu{}^{a(k-2)} ) \\
\delta \Omega_\mu{}^a &=& D_\mu \eta^a - 3b_2 \eta_\mu{}^a, \qquad
\delta B^a = - \eta^a, \qquad 
\delta B^{a(2)} = - \eta^{a(2)} \nonumber
\end{eqnarray}
All equations except the last one are consistent, while for the last
one we obtain:
\begin{equation}
D_{[\mu} B_{\nu]}{}^a = - \Omega_{[\mu,\nu]}{}^a + \frac{5b_2}{6}
e_{[\mu}{}^a B_{\nu]}
\end{equation}
Thus the next equation looks like
\begin{equation}
0 = D_\mu B^{a(2)} + \Omega_\mu{}^{a(2)} - \frac{b_2}{2} [ e_\mu{}^a -
\frac{2}{3} g^{a(2)} B_\mu ] - 2b_3 B_\mu{}^{a(2)}
\end{equation}
Till now all looks exactly as in the spin-3 case, but now gauge
invariance requires that
$$
\delta B^{a(3)} = - \eta^{a(3)}
$$
so the equation for this zero-form also must contain one-forms
and so on. Thus let us consider the chain of equations
($2 \le k \le s-2$):
\begin{eqnarray}
0 &=& D_\mu B^{a(k)} + \Omega_\mu{}^{a(k)} - \frac{(k+2)b_{k+1}}{k} 
B_\mu{}^{a(k)} \nonumber \\
 && - \frac{b_k}{k} [ e_\mu{}^a B^{a(k-1)} - \frac{2}{(2k-1)} g^{a(2)}
B_\mu{}^{a(k-2)} ]  \label{eq1} \\
0 &=& D_\mu B^{a(s-1)} + \Omega_\mu{}^{a(s-1)} - B_\mu{}^{a(s-1)}
\nonumber \\
 && - \frac{b_{s-1}}{(s-1)} [ e_\mu{}^a B^{a(s-2)} - \frac{2}{(2s-3)}
g^{a(2)} B_\mu{}^{a(s-2)} ] \label{eq2}
\end{eqnarray}
where gauge invariance is achieved provided
\begin{equation}
\delta B^{a(k)} = - \eta^{a(k)}, \quad 2 \le k \le s-1, \quad \delta
B^{a(s)} = 0
\end{equation}
All the equations (\ref{eq1}) are consistent, while the consistency of
(\ref{eq2}) gives:
\begin{equation}
D_{[\mu} B_{\nu]}{}^{a(s-1)} = 0
\end{equation}
All zero-forms $B^{a(k)}$ with $k \ge s$ are gauge invariant, so we
obtain all the remaining equations $k \ge s$:
\begin{equation}
0 = D_\mu B^{a(k)} - B_\mu{}^{a(k)} + \frac{(k^2-s^2)}{(2k+1)} \Lambda
[ e_\mu{}^a B^{a(k-1)} - \frac{2}{(2k-1)} B_\mu{}^{a(k-2)} ]
\end{equation}

\subsection{General massive case}

Similarly to the spin-3 case from the very beginning we will use
partially gauge fixed version with separated variables and consider
the subsystem containing fields $\hat{\Omega}$ and $\hat{B}$ only. The
Lagrangian and the whole set of gauge invariant objects are given in
the appendix C and here we begin with the first set of unfolded
equations (with the changed normalization for the zero forms):
\begin{eqnarray}
0 &=& D_{[\mu} \hat{\Omega}_{\nu]}{}^{a(k)} - M_k 
\varepsilon_{[\mu b}{}^a \hat{\Omega}_{\nu]}{}^{a(k-1)b} 
 + \frac{(k+2)b_{k+1}}{k} \hat{\Omega}_{[\mu,\nu]}{}^{a(k)} \nonumber
\\
 && - \frac{b_k}{k} [ e_{[\mu}{}^a \hat{\Omega}_{\nu]}{}^{a(k-1)} +
\frac{2}{(2k-1)} g^{a(2)} \hat{\Omega}_{[\mu,\nu]}{}^{a(k-2)} ]
\nonumber \\
0 &=& D_{[\mu} \hat{\Omega}_{\nu]}{}^a + M_1 \varepsilon_{[\mu}{}^{ab}
\hat{\Omega}_{\nu],b} + 3b_2 \hat{\Omega}_{[\mu,\nu]}{}^a - 2b_1{}^2
e_{[\mu}{}^a \hat{B}_{\nu]} \\
0 &=& D_\mu \hat{B}^a + \hat{\Omega}_\mu{}^a + M_1
\varepsilon_\mu{}^{ab} \hat{B}_b - 3b_2 \hat{B}_\mu{}^a \nonumber
\end{eqnarray}
where parameters $b_k$ and $M_k$ are given in appendix A. We will also
need the gauge transformations:
\begin{eqnarray}
\delta \hat{\Omega}_\mu{}^{a(k)} &=& D_\mu \hat{\eta}^{a(k)} - M_k
\varepsilon_{[\mu b}{}^a \hat{\eta}^{a(k-1)b} - \frac{(k+2)b_{k+1}}{k}
\hat{\eta}_\mu{}^{a(k)} \nonumber \\
 && - \frac{b_k}{k} [ e_\mu{}^a  \hat{\eta}^{a(k-1)} - 
\frac{2}{(2k-1)} g^{a(2)} \eta_\mu{}^{a(k-2)} ] \nonumber \\
\delta \hat{\Omega}_\mu{}^a &=& D_\mu \hat{\eta}^a + M_1 
\varepsilon_\mu{}^{ab} \hat{\eta}_b - 3b_2 \hat{\eta}_\mu{}^a  \\
\delta \hat{B}^a &=& - \hat{\eta}^a, \qquad
\delta B^{a(2)} = - \hat{\eta}^{a(2)} \nonumber
\end{eqnarray}
All equations except the last one are consistent, while for the last
one we obtain:
\begin{equation}
D_{[\mu} \hat{B}_{\nu]}{}^a = - \hat{\Omega}_{[\mu,\nu}{}^a - M_1
\varepsilon_{[\mu}{}^{ab} \hat{B}_{\nu],b} + \frac{5b_2}{6} 
e_{[\mu}{}^a \hat{B}_{\nu]}
\end{equation}
Thus the next equation  has the form:
\begin{equation}
0 = D_\mu \hat{B}^{a(2)} + \hat{\Omega}_\mu{}^{a(2)} - M_2
\varepsilon_{\mu b}{}^a \hat{B}^{ab} - \frac{b_2}{2}
[ e_\mu{}^a \hat{B}^a - \frac{2}{3} g^{a(2)} \hat{B}_\mu ] - 2b_3
\hat{B}_\mu{}^{a(2)}
\end{equation}
where $\delta \hat{B}^{a(3)} = - \hat{\eta}^{a(3)}$ and so on. As in
the partially massless case we proceed with the whole set of equations
\begin{eqnarray}
0 &=& D_\mu \hat{B}^{a(k)} + \hat{\Omega}_\mu{}^{a(k)} - M_k
\varepsilon_{\mu b}{}^a \hat{B}^{a(k-1)b} - \frac{(k+2)b_{k+1}}{k}
\hat{B}_\mu{}^{a(k)} \nonumber \\
 && - \frac{b_k}{k} [ e_\mu{}^a \hat{B}^{a(k-1)} - \frac{2}{(2k-1)}
g^{a(2)} \hat{B}_\mu{}^{a(k-2)} ], \qquad 2 \le k \le s-2 \label{eq3}
\\
0 &=& D_\mu \hat{B}^{a(s-1)} + \hat{\Omega}_\mu{}^{a(s-1)} - M_{s-1}
\varepsilon_{\mu b}{}^a \hat{B}^{a(s-2)b} \nonumber \\
 && - \frac{b_{s-1}}{(s-1)} [ e_\mu{}^a \hat{B}^{a(s-2)} - 
\frac{2}{(2s-3)} g^{a(2)} \hat{B}_\mu{}^{a(s-3)} ] -
\hat{B}_\mu{}^{a(s-1)} \label{eq4}
\end{eqnarray}
where
\begin{equation}
\delta \hat{B}^{a(k)} = - \hat{\eta}^{a(k)}, \quad 2 \le k \le s-1,
\qquad \delta \hat{B}^{a(s)} = 0
\end{equation}
All equations (\ref{eq3}) are consistent, while the consistency for
the (\ref{eq4}) gives:
\begin{equation}
D_{[\mu} \hat{B}_{\nu]}{}^{a(s-1)} = M_{s-1} \varepsilon_{[\mu b}{}^a
\hat{B}_{\nu]}{}^{a(s-2)b}
\end{equation}
Finally taking into account that all $\hat{B}^{a(k)}$ with $k \ge s$
are gauge invariant we obtain all the remaining equations:
\begin{eqnarray}
0 &=& D_\mu \hat{B}^{a(k)} - \hat{B}_\mu{}^{a(k)} - A_k 
\varepsilon_{\mu b}{}^a \hat{B}^{a(k-1)b} \nonumber \\
 && + E_k [ e_\mu{}^a  \hat{B}^{a(k-1)} - \frac{2}{(2k-1)} g^{a(2)}
\hat{B}_\mu{}^{a(k-2)} ]
\end{eqnarray}
\begin{equation}
A_k = \frac{s(s-1)}{k(k+1)} M_{s-1}, \qquad
E_k = \frac{(k^2-s^2)}{(2k+1)} [ \frac{(s-1)^2M_{s-1}{}^2}{k^2} +
\Lambda ]
\end{equation}

\section*{Summary}

Thus we have constructed the unfolded formulation for the massive
bosonic higher spins in three dimensions as well as for their
partially massless limits of the maximal depth. In spite of the number
of features specific for three dimensions, the general picture appears
to be very much like in $d \ge 4$ case \cite{PV10}. Namely, we have a
finite number of zero forms that are not gauge invariant, transform
as the Stueckelberg fields and have equations containing both
one-forms and zero-forms. Besides, we have an infinite number of the
gauge invariant zero-forms those equations contain only zero-forms
themselves. As it was expected, the unfolded formalism in $d=3$ turns
out to be much simpler then the one for $d \ge 4$ so one can hope
that such formalism could be useful for the investigations of possible
interactions.

\section*{Acknowledgments}

Author is grateful to I.~L.~Buchbinder and T.~V.~Snegirev for
collaboration. Work was supported in parts by RFBR grant No.
14-02-01172.

\appendix

\section{Gauge invariant description of massive higher spins}

In three dimensions the frame-like gauge invariant description for the
massive arbitrary spin $s$ \cite{BSZ12a} requires introduction of the
following set of physical and auxiliary fields:
($\Omega_\mu{}^{a(k)}$, $\Phi_\mu{}^{a(k)}$), $1 \le k \le s-1$,
($B^a$, $A_\mu$) and ($\pi^a$, $\varphi$), where $\Omega_\mu{}^{a(k)}$
and $\Phi_\mu{}^{a(k)}$ are symmetric and traceless on their local
indices. The whole Lagrangian consists of the three parts:
$$
{\cal L} = {\cal L}_0 + {\cal L}_1 + {\cal L}_2
$$
\begin{eqnarray}
{\cal L}_0 &=& \sum_{k=1}^{s-1} (-1)^{k+1} [ \frac{k}{2} \eptwo 
\Omega_\mu{}^{ac(k-1)} \Omega_\nu{}^b{}_{c(k-1)} - 
\epthree \Omega_\mu{}^{a(k)} D_\nu \Phi_{\alpha,a(k)} \nonumber \\
 && + \frac{1}{2} B^a B_a - \epthree B_\mu D_\nu A_\alpha - 
\frac{1}{2} \pi^a \pi_a + \pi^\mu D_\mu \varphi \\
{\cal L}_1 &=& \sum_{k=2}^{s-1} (-1)^{k+1} b_k \epthree
[ \frac{(k+1)}{(k-1)} \Omega_{\mu,\nu}{}^{a(k-1)} 
\Phi_{\alpha,a(k-1)} + \Omega_{\mu,a(k-1)} 
\Phi_{\nu,\alpha}{}^{a(k-1)} ] \nonumber \\
 && - b_1 \epthree [ 2 \Omega_{\mu,\nu} A_\alpha - B_\mu 
\Phi_{\nu,\alpha} ] + 2M_1 \pi^\mu A_\mu \\
{\cal L}_2 &=& \sum_{k=1}^{s-1} (-1)^{k+1} \frac{kM_k{}^2}{2} \eptwo 
\Phi_\mu{}^{ac(k-1)} \Phi_\nu{}^b{}_{c(k-1)} \nonumber \\
 && + 2M_1b_1 e^\mu{}_a  \Phi_\mu{}^a \varphi + 3b_1{}^2 \varphi^2
\end{eqnarray}
where ${\cal L}_0$ and ${\cal L}_2$ contain kinetic and mass-like
terms for all fields while ${\cal L}_1$ contains cross-terms gluing
all these fields together. Here:
$$
b_k{}^2 = \frac{(k-1)(s-k)(s+k)}{k(k+1)(2k+1)}
[ m^2 - (s-k-1)(s+k-1) \Lambda ], \qquad k \ge 2
$$
$$
b_1{}^2 = \frac{(s-1)(s+1)}{6} [ m^2 - s(s-2) \Lambda ]
$$
$$
M_k{}^2 = \frac{s^2}{k^2(k+1)^2} [ m^2 - (s-1)^2 \Lambda ]
$$
This Lagrangian is invariant under the following gauge
transformations:
\begin{eqnarray}
\delta \Omega_\mu{}^{a(k)} &=& D_\mu \eta^{a(k)} - 
\frac{(k+2)b_{k+1}}{k} \eta_\mu{}^{a(k)} - M_k{}^2  
\varepsilon_{\mu b}{}^a \eta^{a(k-1)b}  \nonumber \\
 && - \frac{b_k}{k} [ e_\mu{}^a \eta^{a(k-1)} - \frac{2}{(2k-1)}
g^{a(2)} \eta_\mu{}^{a(k-2)} ] \nonumber \\
\delta \Phi_\mu{}^{a(k)} &=& D_\mu \xi^{a(k)} - b_{k+1} 
\xi_\mu{}^{a(k)} - \varepsilon_{\mu b}{}^a \eta^{a(k-1)b} \nonumber \\
 && - \frac{(k+1)b_k}{k(k-1)} [ e_\mu{}^a \xi^{a(k-1)} - 
\frac{2}{2k-1)} g^{a(2)} \xi_\mu{}^{a(k-2)} ] \nonumber \\
\delta \Omega_\mu{}^a &=& D_\mu \eta^a - 3b_2 \eta_\mu{}^a + M_1{}^2
\varepsilon_\mu{}^{ab} \xi_b \nonumber \\
\delta \Phi_\mu{}^a &=& D_\mu \xi^a + \varepsilon_{[\mu}{}^{ab} \eta_b
- b_2 \xi_\mu{}^a + 2b_1 e_\mu{}^a \xi \nonumber \\
\delta B^a &=& - 2b_1 \eta^a, \qquad
\delta A_\mu = D_\mu \xi + b_1 \xi_\mu \nonumber \\
\delta \pi^a &=& 2M_1b_1 \xi^a, \qquad
\delta \varphi = 2M_1 \xi \nonumber
\end{eqnarray}
where the gauge parameters $\eta^{a(k)}$ and $\xi^{a(k)}$ are also
symmetric and traceless.

\section{Partially massless limit}

From the explicit formulas given in the previous appendix, one can
see that in de-Sitter space $\Lambda > 0$ there exists a number of
special mass values where one of the parameters $b_l = 0$. In this
case the whole system decomposes into two independent subsystems. The
first one with the fields ($\Omega_\mu{}^{a(k)}$, 
$\Phi_\mu{}^{a(k)}$), $l \le k \le s-1$, describes the so called
partially massless theory, while the remaining fields gives massive
spin-$l$. In three dimensions most of such partially massless fields
do not have any physical degrees of freedom and do not require
introduction of any zero-forms. The only case with one physical
degree of freedom corresponds to
\begin{equation}
M_1 = 0 \quad \Rightarrow \quad m^2 = (s-1)^2 \Lambda
\end{equation}
when spin-0 decouples. Note that in this case all $M_k = 0$ so that
the Lagrangian and gauge transformations are greatly simplified:
\begin{eqnarray}
{\cal L}_0 &=& \sum_{k=1}^{s-1} (-1)^{k+1} [ \frac{k}{2} \eptwo 
\Omega_\mu{}^{ac(k-1)} \Omega_\nu{}^b{}_{c(k-1)} - \epthree 
\Omega_\mu{}^{a(k)} D_\nu  \Phi_{\alpha,a(k)} ] \nonumber \\
&& + \sum_{k=2}^{s-1} (-1)^{k+1} b_k \epthree [ \frac{(k+1)}{(k-1)}
\Omega_{\mu,\nu}{}^{a(k-1)} \Phi_{\alpha,a(k-1)} + 
\Omega_{\mu,a(k-1)}  \Phi_{\nu,\alpha}{}^{a(k-1)} ] \nonumber \\
 && + \frac{1}{2} B^a B_a - \epthree B_\mu D_\nu A_\alpha + b_1
\varepsilon^{\mu\nu\alpha} [ - 2 \Omega_{\mu,\nu} A_\alpha + B_\mu
\Phi_{\nu,\alpha} ]  
\end{eqnarray}
where now
\begin{equation}
b_k{}^2 = \frac{k(k-1)(s^2-k^2)}{(k+1)(2k+1)} \Lambda, \qquad
b_1{}^2 = \frac{(s^2-1)}{6} \Lambda \label{b_param}
\end{equation}
\begin{eqnarray}
\delta \Omega_\mu{}^{a(k)} &=& D_\mu \eta^{a(k)} - 
\frac{(k+2)b_{k+1}}{k} \eta_\mu{}^{a(k)} - \frac{b_k}{k} [ e_\mu{}^a
\eta^{a(k-1)} - Tr ] \nonumber \\
\delta \Phi_\mu{}^{a(k)} &=& D_\mu \xi^{a(k)} - 
\varepsilon_{\mu b}{}^a \eta^{a(k-1)b} - b_{k+1} \xi_\mu{}^{a(k)} -
\frac{(k+1)b_k}{k(k-1)} [ e_\mu{}^a \xi^{a(k-1)} - Tr ] \nonumber \\
\delta \Omega_\mu{}^a &=& D_\mu \eta^a - 3b_2 \eta_\mu{}^a  \\
\delta \Phi_\mu{}^a &=& D_\mu \xi^a - b_2 \xi_\mu{}^a + 
\varepsilon_\mu{}^{ab} \eta_b + 2b_1 e_\mu{}^a \xi \nonumber \\
\delta B^a &=& - 2b_1 \eta^a, \qquad
\delta A_\mu = D_\mu \xi + b_1 \xi_\mu \nonumber
\end{eqnarray}
As usual in the frame-like gauge invariant formalism, for each field
(both physical as well as auxiliary one) one can construct
corresponding gauge invariant object:
\begin{eqnarray}
{\cal F}_{\mu\nu}{}^{a(k)} &=& D_{[\mu} \Omega_{\nu]}{}^{a(k)} +
\frac{(k+2)b_{k+1}}{k} \Omega_{[\mu,\nu]}{}^{a(k)} - \frac{b_k}{k}
[ e_{[\mu}{}^a \Omega_{\nu]}{}^{a(k-1)} - Tr ] \nonumber \\
{\cal T}_{\mu\nu}{}^{a(k)} &=& D_{[\mu} \Phi_{\nu]}{}^{a(k)} -
\varepsilon_{[\mu b}{}^a \Omega_{\nu]}{}^{a(k-1)b} + b_{k+1}
\Phi_{[\mu,\nu]}{}^{a(k)} \nonumber \\
 && - \frac{(k+1)b_k}{k(k-1)} [ e_{[\mu}{}^a
\Phi_{\nu]}{}^{a(k-1)} - Tr ] \nonumber \\
{\cal R}_{\mu\nu}{}^a &=& D_{[\mu} \Omega_{\nu]}{}^a + 3b_2
\Omega_{[\mu,\nu]}{}^a - b_1 e_{[\mu}{}^a B_{\nu]} \nonumber \\
{\cal T}_{\mu\nu}{}^a &=& D_{[\mu} \Phi_{\nu]}{}^a +
\varepsilon_{[\mu}{}^{ab} \Omega_{\nu],b} + b_2 \Phi_{[\mu,\nu]}{}^a
+ 2b_1 e_{[\mu}{}^a A_{\nu]} \\
{\cal B}_\mu{}^a &=& D_\mu B^a + 2b_1 \Omega_\mu{}^a - B_\mu{}^a
\nonumber \\
{\cal A}_{\mu\nu} &=& D_{[\mu} A_{\nu]} - 
\varepsilon_{\mu\nu a} B^a - b_1 \Phi_{[\mu,\nu]} \nonumber
\end{eqnarray}
where the gauge invariance requires the introduction of the zero-form 
$B^{ab}$ which does not enter the free Lagrangian and transforms
non-trivially under the $\eta^{ab}$ transformation:
\begin{equation}
\delta B^{ab} = - 6b_2b_1 \eta^{ab}
\end{equation}

\section{Partial gauge fixing}

As is well known \cite{AT86,Wit88} (see also \cite{Gom13} and
references therein) in the frame-like formalism for the massless
higher spin fields in the three dimensional anti-de Sitter space one
can introduce combinations of physical and auxiliary fields such that
the whole theory (not only the free theory but an interacting one as
well) decomposes into two independent subsystems. It was shown in
\cite{BSZ12a} that such separation works for the massive higher spins
as well provided one uses a partial gauge fixing to remove scalar
field. Moreover, as the frame-like gauge invariant description for
massive fields itself, such mechanism works not only in anti-de Sitter
space but in Minkowski and de Sitter spaces provided $m^2 > (s-1)^2
\Lambda$.

Let us partially gauge fix the general massive theory described in
appendix A by setting the gauge $\varphi = 0$, solve the constraint 
$A_\mu = \pi_\mu/(2M_1)$ and re-scale $\pi^a \Rightarrow
2M_1\pi^a$ (taking into account that $\pi^a$ will play now the role of
physical field and not that of the auxiliary one). Resulting
Lagrangian takes the form:
\begin{eqnarray}
{\cal L} &=& \sum_{k=1}^{s-1} (-1)^{k+1} [ \frac{k}{2} \eptwo 
\Omega_\mu{}^{ac(k-1)} \Omega_\nu{}^b{}_{c(k-1)} - 
\epthree \Omega_\mu{}^{a(k)} D_\nu  \Phi_{\alpha,a(k)} \nonumber \\
 && + \sum_{k=2}^{s-1} (-1)^{k+1} b_k \epthree
[ \frac{(k+1)}{(k-1)} \Omega_{\mu,\nu}{}^{a(k-1)} 
\Phi_{\alpha,a(k-1)} + \Omega_{\mu,a(k-1)} 
\Phi_{\nu,\alpha}{}^{a(k-1)} ] \nonumber \\
 && + \frac{1}{2} B^a B_a - \epthree B_\mu D_\nu  \pi_\alpha - b_1
\epthree [ 2 \Omega_{\mu,\nu}  \pi_\alpha - B_\mu \Phi_{\nu,\alpha} ]
\nonumber \\
 && + \sum_{k=1}^{s-1} (-1)^{k+1} \frac{kM_k{}^2}{2} \eptwo 
\Phi_\mu{}^{ac(k-1)} \Phi_\nu{}^b{}_{c(k-1)} + 2M_1{}^2 \pi^a
\pi_a
\end{eqnarray}
Let us introduce new variables:
\begin{equation}
\hat{\Omega}_\mu{}^{a(k)} = \Omega_\mu{}^{a(k)} + M_k 
\Phi_\mu{}^{a(k)}, \qquad
\hat{\Phi}_\mu{}^{a(k)} = \Omega_\mu{}^{a(k)} - M_k 
\Phi_\mu{}^{a(k)} 
\end{equation}
\begin{equation}
\hat{B}^a = B^a - 2M_1 \pi^a, \qquad
\hat{\pi}^a = B^a + 2M_1 \pi^a
\end{equation}
Then the whole Lagrangian can be rewritten as:
$$
{\cal L} = {\cal L}(\hat{\Omega},\hat{B}) - 
{\cal L}(\hat{\Phi},\hat{\pi})
$$
where, for example,
\begin{eqnarray}
{\cal L}(\hat{\Omega},\hat{B}) &=& \sum_{k=1}^{s-1} 
\frac{(-1)^{k+1}}{4M_k} [ kM_k \eptwo \hat{\Omega}_\mu{}^{ac(k-1)}
\hat{\Omega}_\nu{}^b{}_{c(k-1)} - \epthree \hat{\Omega}_\mu{}^{a(k)} 
D_\nu \hat{\Omega}_{\alpha,a(k)} ] \nonumber \\
 && + \frac{1}{4M_1} [ M_1 \hat{B}^a \hat{B}_a +
\epthree \hat{B}_\mu D_\nu \hat{B}_\alpha ] \nonumber \\
 && + \sum_{k=2}^{s-1} \frac{(-1)^{k+1}b_k}{2M_k} \epthree
\hat{\Omega}_{\mu,\nu}{}^{a(k-1)} \hat{\Omega}_{\alpha,a(k-1)} +
\frac{b_1}{2M_1} \epthree \hat{\Omega}_{\mu,\nu} \hat{B}_\alpha
\end{eqnarray}
This Lagrangian is invariant under the following gauge
transformations;
\begin{eqnarray}
\delta \hat{\Omega}_\mu{}^{a(k)} &=& D_\mu \hat{\eta}^{a(k)} - M_k
\varepsilon_{\mu b}{}^a \hat{\eta}^{a(k-1)b} - \frac{(k+2)b_{k+1}}{k}
\hat{\eta}_\mu{}^{a(k)} \nonumber \\
 && - \frac{b_k}{k} [ e_\mu{}^a  \hat{\eta}^{a(k-1)} - Tr ] \nonumber
\\
\delta \hat{\Omega}_\mu{}^a &=& D_\mu \hat{\eta}^a + M_1 
\varepsilon_\mu{}^{ab} \hat{\eta}_b - 3b_2 \hat{\eta}_\mu{}^a \\
\delta \hat{B}^a &=& - 2b_1 \hat{\eta}^a \nonumber
\end{eqnarray}
where
\begin{equation}
\hat{\eta}^{a(k)} = \eta^{a(k)} + M_k \xi^{a(k)}
\end{equation}
Moreover, for each field we can still construct corresponding gauge
invariant object:
\begin{eqnarray}
\hat{\cal F}_{\mu\nu}{}^{a(k)} &=& D_{[\mu} 
\hat{\Omega}_{\nu]}{}^{a(k)} - M_k \varepsilon_{[\mu b}{}^a
\hat{\Omega}_{\nu]}{}^{a(k-1)b} + \frac{(k+2)b_{k+1}}{k}
\hat{\Omega}_{[\mu,\nu]}{}^{a(k)} \nonumber \\
 && - \frac{b_k}{k} [ e_{[\mu}{}^a \hat{\Omega}_{\nu]}{}^{a(k-1)} - Tr
] \nonumber \\
\hat{\cal F}_{\mu\nu}{}^a &=& D_{[\mu} \hat{\Omega}_{\nu]}{}^a + M_1
\varepsilon_{[\mu}{}^{ab} \hat{\Omega}_{\nu],b} + 3b_2
\hat{\Omega}_{[\mu,\nu]}{}^a - b_1 e_{[\mu}{}^a \hat{B}_{\nu]} \\
\hat{\cal B}_\mu{}^a &=& D_\mu \hat{B}^a + 2b_1 \hat{\Omega}_\mu{}^a
+ M_1 \varepsilon_\mu{}^{ab} \hat{B}_b - \hat{B}_\mu{}^a \nonumber
\end{eqnarray}
where, similarly to the partially massless case, gauge invariance
requires introduction of the zero-form $\hat{B}^{ab}$ such that
\begin{equation}
\delta \hat{B}^{ab} = - 6b_1b_2 \hat{\eta}^{ab}
\end{equation}

\end{document}